\RequirePackage{rotating}
\RequirePackage{fix-cm}
\documentclass[smallextended]{svjour3}       
\smartqed  
\usepackage[table]{xcolor}
\usepackage{wrapfig}
\input{insbox}
\usepackage{soul}
\usepackage{graphicx}
\usepackage[misc,geometry]{ifsym}
\usepackage{ulem}
\usepackage{adjustbox}
\usepackage{framed}
\usepackage[utf8]{inputenc}
\usepackage{booktabs}
\usepackage{tcolorbox}
\usepackage{amsmath}
\usepackage{amssymb}
\usepackage{paralist}
\usepackage{multirow}
\usepackage{xspace}
\usepackage{color}
\usepackage{xcolor}
\usepackage[graphicx]{realboxes}
\usepackage{ifthen}
\usepackage{url}
\usepackage{fancybox}
\usepackage{enumitem}
\usepackage{listings}
\usepackage{balance}
\usepackage{ifthen}
\usepackage{natbib}
\usepackage{graphicx}
\usepackage{amsmath}
\usepackage[linesnumbered,boxruled]{algorithm2e}
\usepackage{algorithm2e}
\usepackage{tablefootnote}
\usepackage{comment}
\usepackage{float}
\usepackage{subfigure}
\usepackage{algorithmic}
\usepackage{dirtytalk}
\usepackage{tcolorbox}
\definecolor{mygray}{gray}{0.6}
\usepackage[flushleft]{threeparttable}
\usepackage{scalerel,graphicx,xparse}

\NewDocumentCommand\emojiThumbsUp{}{\scalerel*{\includegraphics{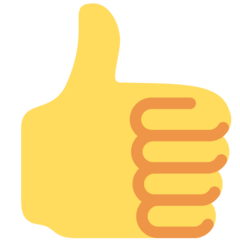}}{X}}
\NewDocumentCommand\emojiThumbsDown{}{\scalerel*{\includegraphics{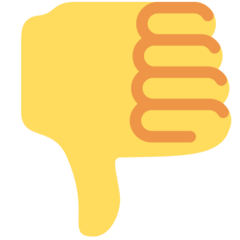}}{X}}
\NewDocumentCommand\emojiTada{}{\scalerel*{\includegraphics{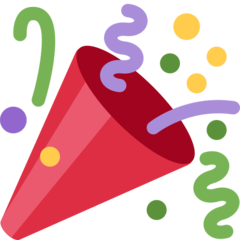}}{X}}
\NewDocumentCommand\emojiHeart{}{\scalerel*{\includegraphics{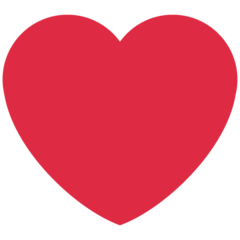}}{X}}
\NewDocumentCommand\emojiRocket{}{\scalerel*{\includegraphics{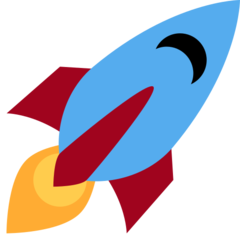}}{X}}
\NewDocumentCommand\emojiConfused{}{\scalerel*{\includegraphics{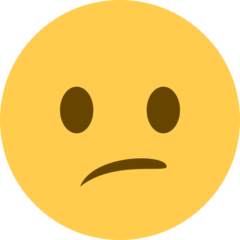}}{X}}
\NewDocumentCommand\emojiEyes{}{\scalerel*{\includegraphics{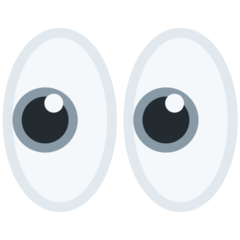}}{X}}
\NewDocumentCommand\emojiGrinning{}{\scalerel*{\includegraphics{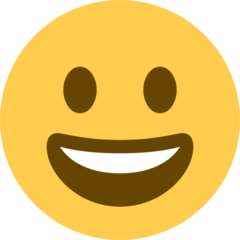}}{X}}

\newlength\WIDTHOFBAR
\def\mybar#1{
  {\color{gray}\rule{#1cm}{8pt}}}
  
\def\mybar#1{
  #1s & {\color{grey}\rule{#1cm}{8pt}}}

\definecolor{chestnut}{rgb}{0.8, 0.36, 0.36}

\definecolor{chestnut}{rgb}{0.8, 0.36, 0.36}

\usepackage[colorlinks = true,
            linkcolor = black,
            urlcolor  = black,
            citecolor = black,
            anchorcolor = black]{hyperref}

\DeclareGraphicsExtensions{.pdf,.jpeg,.png}

\def\mybar#1{
  {\color{gray}\rule{#1cm}{8pt}}}

\begin{document}


\title{More Than React: Investigating The Role of Emoji
Reaction in GitHub Pull Requests}
\author{Dong Wang \Letter\and Tao Xiao \and Teyon Son \and Raula Gaikovina Kula  \and Takashi Ishio \and Yasutaka Kamei \and Kenichi Matsumoto 
}

\institute{
    \Letter~Corresponding author - Dong Wang
    \at Kyushu University, Japan.\\
    \email{d.wang@ait.kyushu-u.ac.jp}
     \and
    Tao Xiao, Teyon Son, Raula Gaikovina Kula, Takashi Ishio, Kenichi Matsumoto
    \at Nara Institute of Science and Technology, Japan.\\
    \email{\{tao.xiao.ts2,son.teyon.sr7,raula-k,ishio,matumoto\}@is.naist.jp}
    \and Yasutaka Kamei
    \at Kyushu University, Japan.
    \email{
kamei@ait.kyushu-u.ac.jp}
}

\date{Received: date / Accepted: date}
\maketitle
\begin{abstract}
Open source software development has become more social and collaborative, evident GitHub. 
Since 2016, GitHub started to support more informal methods such as emoji reactions, with the goal to reduce commenting noise when reviewing any code changes to a repository.
From a code review context, the extent to which emoji reactions facilitate a more efficient review process is unknown.
We conduct an empirical study to mine 1,850 active repositories across seven popular languages to analyze 365,811 Pull Requests (PRs) for their emoji reactions against the review time, first-time contributors, comment intentions, and the consistency of the sentiments.
Answering these four research perspectives, we first find that the number of emoji reactions has a significant correlation with the review time. 
Second, our results show that a PR submitted by a first-time contributor is less likely to receive emoji reactions.  
Third, the results reveal that the comments with an intention of \textit{information giving}, are more likely to receive an emoji reaction.
Fourth, we observe that only a small proportion of sentiments are not consistent between comments and emoji reactions, i.e., with 11.8\% of instances being identified. 
In these cases, the prevalent reason is when reviewers cheer up authors that admit to a mistake, i.e., \textit{acknowledge a mistake}.
Apart from reducing commenting noise, our work suggests that emoji reactions play a positive role in facilitating collaborative communication during the review process.


\end{abstract}

\section{Introduction}
\label{intro}
In the past few years, open source software development has become more social and collaborative.
Known as social coding, open source development promotes formal and informal collaboration by empowering the exchange of knowledge between developers~\citep{dabbish2012social}. 
GitHub, one of the most popular social coding platforms, attracts more than 72 million developers collaborating across 233 million repositories.\footnote{\url{https://github.com/search, 2021}}
Since 2016, GitHub introduced a new social function called ``\textit{reaction}'' for developers to quickly express their feeling in an issue report and a pull request (PR).
Especially for discussing a PR, we find that\footnote{\url{https://tinyurl.com/3rpdr6dp}}:
\begin{quote}
    \textit{``In many cases, especially on popular projects, the result is a long thread full of emoji and not much content, which makes it difficult to have a discussion. With reactions, you can now reduce the noise in these threads'' - GitHub}
\end{quote}

\begin{figure*}[t]
    \centering
    \subfigure[Example of emoji reaction reduce commenting noise.\label{fig:em}]{\includegraphics[width=0.8\textwidth]{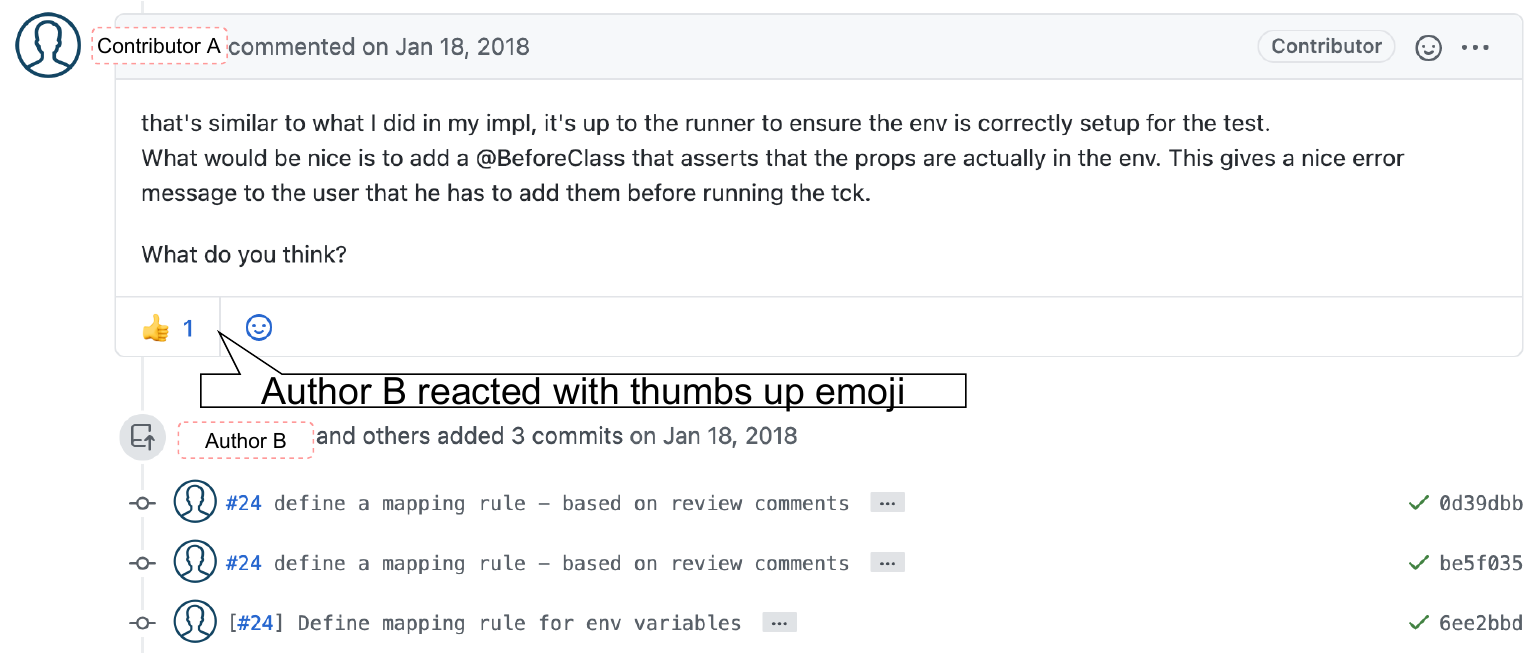}}
        \subfigure[Example of emoji reaction does not reduce commenting noise.\label{fig:moti}]{\includegraphics[width=0.8\textwidth]{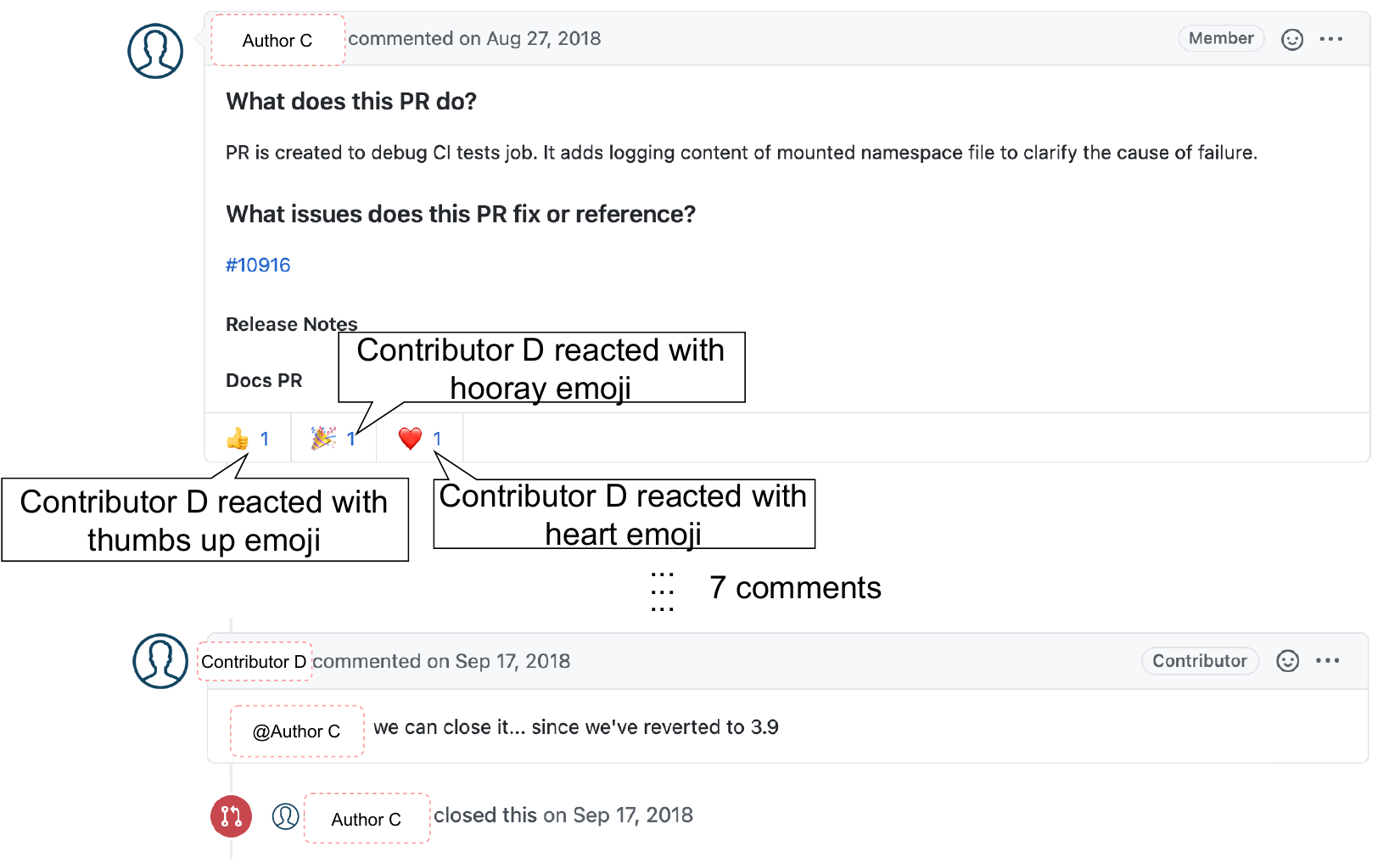}}
    \caption{ Examples of emoji reactions used in GitHub.}
    \label{fig:examples}
\end{figure*}

In the context of code review, we assume that a thread full of emoji reactions may also have ulterior intentions (e.g., confusion or conflicts) during the code review process.
For instance, \citet{confusion_saner_2019} pointed out that confusion delays the merge decision decreases review quality, and results in additional discussions.
\citet{hirao2020code} found that patches can receive both positive and negative scores due to the disagreement between reviewers, which leads to conflicts in the review process.
Figure \ref{fig:examples} depicts two typical cases where the emoji reactions occur.
Figure \ref{fig:em} shows the case where the reaction does reduce unnecessary commenting in the thread.
The example illustrates how \texttt{Author B} reduces the commenting by simply reacting with a quick expression of approval through \texttt{THUMBS UP} \emojiThumbsUp.
In contrast, as shown in Figure~\ref{fig:moti}, there exists a case where the emoji usage has an ulterior intention and does not reduce comments in the discussion thread. 
In detail, \texttt{Contributor D} uses three positive emoji reactions (\texttt{THUMBS UP} \emojiThumbsUp, \texttt{HOORAY} \emojiTada, and \texttt{HEART} \emojiHeart) to represent the appreciation to this PR.
Later the contributor goes on to provide detailed comments on the PR. 
We posit that the intention of the emoji reaction was to express appreciation for the PR, and did not reduce the amount of commenting in the threads.

As per our registered report~\citep{son2021more}, positive emoji reactions are been widely used in PRs and do not always reduce the commenting noise.
We now execute the study protocols to address four research questions, using a statistical sample of 1,850 repositories with 365,811 PRs across seven popular languages.
The goal of the study is to investigate the role of emoji reactions in the context of code review from different perspectives. 
RQ1 is from the view of the review process, while RQ2 is from the human (i.e., contributor) perspective.
Finally, RQ3 and RQ4 take a deeper analysis of the context of the comments between the review team.
We now present the details of each research question below:
\begin{itemize}
\item \textbf{RQ1: \textit{Does the emoji reaction used in the review discussion correlate with review time?}}\\
\textbf{\emph{Motivation.}}
Prior studies \citep{Olga_2016, Shopify_2018} have widely analyzed the impact of technical and non-technical factors on the review process (e.g., review outcome, review time). 
However, little is known about whether or not the emoji reaction can be correlated with the review time.
It is possible that emoji reaction may shorten the review time, as it could reduce the noise during the review discussions.
Thus, our motivation for the first research question is to explore the correlation between the emoji reaction used in the review discussion and review time.
\item\textbf{RQ2: \textit{Does a PR submitted by a first-time contributor receive more emoji reactions?}}\\
\textbf{\emph{Motivation.}} 
We find that the emoji reaction might be used to express appreciation for submitting a PR.
Our motivation for this research question is to understand if contributors that have never submitted to the project before receive more emoji reactions.
Furthermore, answering this research question will provide insights into a potent ulterior motive for an emoji reaction.
Our hypothesis is that \textit{H1: PRs submitted by first-time contributors receive more emoji reactions.} 
We assume that existing contributors could express positive feelings to attract newcomers to the project.
\item\textbf{RQ3: \textit{What is the relationship between the intention of comments and their emoji reaction?}}\\
\textbf{\emph{Motivation.}} As shown in Figure 1,  emoji reactions may not always reduce the commenting noise.
Hence, our motivation for the third research question is to explore the relationship between the intention of comments and their emoji reactions. 
Our hypothesis is that \textit{(H2): There
is a significant relationship between comment intentions and emoji reaction kinds.}
We assume that a specific comment intention may explain the ulterior purpose of reacting with an emoji reaction.
\item\textbf{{RQ4: \textit{Is emoji reaction consistent with comment sentiment?}}}\\
\textbf{\emph{Motivation.}} The RQ3 results reveal that specific sentiments of the emoji (i.e., \texttt{THUMBS UP} \emojiThumbsUp) are widely used in PRs.
Inspired by this, our motivation for this research question is to investigate whether there is any inconsistency between sentiments of the comments and sentiments of the emoji reactions.
Furthermore, we manually check the reasons why inconsistency happened. We believe answering RQ4 would help newcomers better understand the emoji usage in the PR discussion.
Our hypothesis is that \textit{(H3): There is a significant relationship between comment sentiments and emoji reaction sentiments.}
We assume that a specific comment sentiment may explain the ulterior purpose of reacting with an emoji.
\end{itemize}

Our results of each research question are summarized as follows.
For \textit{RQ1}, our regression model shows that the number of emoji reactions has a significant correlation with the review time and furthermore PRs with emoji reactions overall tend to take a longer review time than PRs with no emoji reactions.
For \textit{RQ2}, the quantitative results show that a PR submitted by a first-time contributor is less likely to receive emoji reactions, being 10.4\% of PRs.
For \textit{RQ3}, we observe that the PR comments with the intention of information giving are more likely to receive emoji reactions. 
Moreover, the positive THUMBS UP (67.2\% on average) is widely used, while the negative emoji reactions are rarely used (0.5\% to 1.5\%).
For \textit{RQ4}, our results show that Positive–Negative pair-wise of sentiment inconsistency accounts for 11.8\% and the most frequent inconsistency reason is to acknowledge a mistake.
These results suggest that the usage of emoji reactions might be a sign of an already positive environment and could have the potential to reduce toxicity.
Our study demonstrates the crucial role that emoji reactions play in facilitating collaborative communication during the review process.

The remainder of this paper is organized as follows.
Section 2 describes the dataset preparation.
Section 3 presents the results of our empirical study, including the approaches for research questions, while
Section 4 discusses the findings and their implications.
Section 5 discloses the threats to validity and Section 6 presents the related work.
Section 7 discusses the deviations between the execution and the registered report.
Finally, we conclude the paper in Section 8.

\section{Dataset Preparation}
In this section, we present our dataset that was collected for our experiments.

\paragraph{\textbf{Studied Repositories.}} We expand on our studied dataset from the active software development repositories shared by \citet{hata}.
Specifically, each repository has more than 500 commits and at least 100 commits during the most active two-year period.
In total, 25,925 repositories were contained across seven popular languages (i.e., C, C++, Java, JavaScript, Python, PHP, and Ruby).
Since we focus on the pull request, we did further filtering to collect the repository candidates that actively perform code review activities. 
To do so, we automatically examined repositories whether they have at least 100 PRs after the GitHub emoji reaction was introduced to the community (Mar. 10, 2016), relying on GitHub GraphQL API.\footnote{\url{https://docs.github.com/en/graphql}}
After this, 6,695 repositories met the selection criteria.

However, it is impossible to retrieve the PR metadata for all these repositories due to the limited time-frame and the restriction of GitHub API downloading.
We then draw a representative repository sample, taking the seven popular languages into account.
With a confidence level of 95\% and a confidence interval of 5,\footnote{\url{https://www.surveysystem.com/sscalc.htm}} 1,850 representative repositories were finally selected as shown in Table \ref{Data_summary} (i.e., 217 C repositories, 258 C++ repositories, 290 Java repositories, 310 JavaScript repositories, 246 PHP repositories, 300 Python repositories, and 229 Ruby repositories.)

\begin{table}[b]
\caption{Summary of Studied Dataset Statistics. E.R. refers to Emoji Reaction.}
\label{Data_summary}
\resizebox{\textwidth}{!}{
\begin{tabular}{lrrrrr}
\toprule
           & \# Repos. & \# PRs & \# PRs with E.R. & \# PR Comments & \# PR Comments with E.R. \\ \midrule
C          & 217                & 50,847    & 7,330 (14.4\%)                          &      177,540             &               11,377 (6.4\%)                         \\
C++        & 258                &   70,877        &    10,920 (15.4\%)                            &       253,062            &       18,804 (7.4\%)                                \\
Java       & 290                &    69,473      &    6,853 (9.9\%)                             &        219,477          &     11,102 (5.1\%)                                   \\
JavaScript & 310                &   36,135        &  4,467 (12.4\%)                              &     82,623            &     6,591 (8.0\%)                                   \\
Python     & 300                &    53,996       &      5,893 (10.9\%)                          &   158,535                &               8,967 (5.7\%)                       \\
PHP        & 246                &     49,680      &   9,165 (18.4\%)                             &   207,345                &          15,871 (7.7\%)                             \\
Ruby       & 229                &   34,803        &       4,832 (13.9\%)                         &  89,996                 &        7,906 (8.8\%)                              \\ \midrule
\rowcolor{gray!30}
Total &   1,850       &    365,811   &   49,460 (13.5\%)   &    1,188,578   &    80,618 (6.8\%)   \\
\bottomrule
\end{tabular}}
\end{table}

\paragraph{\textbf{Data Collection and Cleaning.}}
For the selected 1,850 repositories, we then used GraphQL API to retrieve all PRs that were submitted between January 2020 and April 2022.
We argue that the more recent PRs, the more emojis will be reacted to PR comments.
Sufficient metadata was collected including PR title, PR status, PR author, created time, closed time, and comments.
For each comment in a PR, we collected its commenters and emoji reactions.
In all, 554,112 PRs with 1,697,305 comments were retrieved from these 1,850 repositories across seven languages.

We then did two filters to ensure the quality of the studied PRs.
First, we excluded PRs that were labeled as open status,
since we can not calculate the review time of these PRs.
The second filter is to exclude the PRs that were submitted by bots.
To do so, we referred to the work of \citet{golzadeh2022accuracy}, which systematically compares the performance of the existing bot detection techniques, and we leveraged the combination of two bot detection techniques with the highest precision known as ``bot'' suffix and list of bots.
``bot'' suffix refers to the technique that relies on the presence of the string ``bot'' at the end of the author’s name, which has notably been used by other researches \citep{dey2020detecting, saadat2021bots}. 
The list of bots denotes the technique that relies on a predefined list of ground-truth bots manually identified by \citet{golzadeh2021ground}.
After the cleaning, 365,811 PRs were left with 1,188,578 PR comments.
As shown in Table \ref{Data_summary}, 13.5\% of PRs across seven languages contain at least one emoji reaction.
80,618 PR comments are reacted with emojis, accounting for 6.8\%.

\section{Empirical Results}
In this section, we present the results for each of our research questions as well as their motivations and approaches.

\subsection{Review Time and Emoji Reactions (RQ1)}
\label{section:4.1}

\noindent
\textbf{\emph{Approach.}}
To answer RQ1, we perform a statistical analysis to investigate the correlation between the PR review time and the emoji reaction kinds, using a non-linear regression model.
Align with the prior work~\citep{WangEMSE2021, wang2021can, Shopify_2018}, several confounding factors are taken into account. 
Similarly, the goal of our statistical analysis is not to predict the review time but to understand the associations between the emoji reaction and the review time.
The review time is defined as the time interval between pull request creation date and closed date, in hours, referring to the work of \citet{maddila2019predicting}.

\begin{table}[t]
\centering
\caption{The studied explanatory variables in RQ1.}
\begin{tabular}{l|p{7cm}}
\toprule
\rowcolor{gray!30}
Confounding variables       & Description                                      \\ \midrule
\# Added lines                     & The number of added LOC by a PR.                                       \\
\# Deleted lines                   & The number of deleted LOC by a PR.                                      \\
PR size               & The total number of added and deleted LOC by a PR.                   \\
Purpose              & The purpose of a PR, i.e., bug, document, feature.                  \\
Language                & The repository language that a PR belongs to.                     \\
\# Files                 & The number of files changed by a PR.                     \\
\# Commits & The number of commits involved in a PR. \\
Description length            & The length of a PR description.                     \\
PR author experience  & The number of prior PRs that were submitted by the PR author.     \\
\# Comments              & The number of comments left on a PR.                      \\
\# Author comments       & The number of comments left by the PR author.         \\
\# Reviewer comments     & The number of comments left by the reviewers who participate in the discussion.             \\
\# Reviewers & The number of developers who participate in the discussion.              \\\midrule
\rowcolor{gray!30}
Emoji reaction variables & Description \\\midrule
With emoji reaction        & Whether or not a PR contains any emoji reactions (binary). \\
\# Emoji reactions       & The number of emoji reactions in a PR. \\\bottomrule
\end{tabular}
\label{table:variables}
\end{table}

\textit{Explanatory Variables.} Table \ref{table:variables} presents the 14
studied explanatory variables that are used in our non-linear regression model.
Since we investigate the effect of the emoji reaction, we introduce two related variables: \texttt{With emoji reaction} (Whether or not a PR contains any emoji reactions, binary) and \texttt{\# Emoji reactions} (The number of emoji reactions in a PR).
Previous research suggests a group of different metrics that can affect code review time~\citep{Shopify_2018, WangEMSE2021} and we select 12 variables that are governed by our ability to accurately calculate their value from the data.
For the PR purpose, similar to the prior work~\citep{EMSE2017_Pick, shane2014msr}, a PR is classified as documentation if the description contains `doc', `copyright', or `license' words, while a PR is classified as bug fixing if the description contains `fix', `bug', or `defect' words.
The rest of the PRs are classified as feature introduction.


\textit{Model Construction.} To reduce the potential threat resulting from the imbalanced data, we randomly select 49,460 PRs without any emoji reactions, which are equal to the number of PRs with emoji reactions (as shown in Table \ref{Data_summary}). 
That is, a total of 98,920 PRs are used in our model construction. To analyze the association between the emoji reaction and review time, we adopt the Ordinary Least Squares (OLS) multiple regression model.
This regression model allows us to fit the nonlinear relationship between the dependent variables and the explanatory variables.
We carefully follow the construction approach provided by \citet{MC1_refer} and \citet{EMSE_quality2016}, consisting of five steps.
In step \textit{(I) Estimating budget for degrees of freedom,} 
as suggested by \citet{MC1_refer}, we spend no more than $\frac{n}{15}$ degrees of freedom in our OLS model, where n refers to the number of studied PRs in the dataset.
\textit{In step (II) Normality adjustment}, we analyze whether the distribution of review time is skewed using the \texttt{skewness} and \texttt{kurtosis} function of the \texttt{moments} R package, since OLS expects that the dependent variables are normally distributed.
If the review time is skewed, we use a log transformation to lessen the skew so as to better fit OLS~\citep{EMSE_quality2016}.
\textit{In step (III) Correlation and redundancy analysis}, similar to the prior work, we use the Spearman rank correlation ($\rho$) to assess the correlation between each pair of variables, since highly correlated explanatory variables could interfere with each other and further lead to spurious conclusions. 
We repeat the process until the Spearman correlation coefficient values of all pairs of variables are less than 0.7.
In addition, to ensure that each studied variable provides a unique signal, we use the \texttt{redun} function of the \texttt{rms} R package to detect redundant variables and remove them from the model.
In step \textit{(IV) Allocating degrees of freedom}, we rely on the \texttt{spearman2} function of the \texttt{rms} R package to calculate the Spearman multiple $\rho^2$ between the explanatory and dependent variables, and effectively allocate degrees of freedom to the remaining variables.
To avoid the over-fitting issue, we only allocate three to five degrees of freedom to those variables with high $\rho^2$ values.
In step \textit{(V) Fitting OLS models}, similar to the work~\citep{EMSE_quality2016}, we use restricted cubic splines to fit our modeled dataset.
We  assign the allocated degrees of freedom to each explanatory variable, using the \texttt{rcs} function of the \texttt{rms} R package.
Last, we adopt the \texttt{ols} function to construct the model.

\textit{Model Analysis.}
After the model construction, we analyze the model to assess its goodness and examine the relationship between the review time and the emoji reaction.
Similar to the prior work~\citep{EMSE_quality2016,WangEMSE2021}, we analyze the model using the following three steps: \textit{(I) Assessing model stability}, \textit{(II) Estimating the power of explanatory variables}, and \textit{(III) Examining relationship}.
In step (I), we use an adjusted $R^2$ value~\citep{hastie2009elements} to evaluate our model.
In order to avoid the overfitted model, we apply the bootstrap validation approach to estimate the optimism of the adjusted $R^2$.
Finally, we subtract the average $R^2$ optimism from the initial adjusted $R^2$ value to obtain the optimism-reduced adjusted $R^2$.
In step (II), we estimate the power of explanatory variables and their corresponding significance, using Wald $\chi^2$ maximum likelihood tests provided by the \texttt{anova} function. 
In step (III), we examine the direction of the relationship between the explanatory variables (especially emoji reaction related variables) and the review time, using the \texttt{Predict} function of the \texttt{rms} package.


\begin{table}[]
\centering
\caption{PR review time model statistics (RQ1). Among studied emoji reaction variables, \# Emoji Reactions have a significant correlation with PR review time.}
\label{RQ1_Statistics}
\resizebox{.7\textwidth}{!}{
\begin{threeparttable}
\begin{tabular}{ll|rr}
\multicolumn{4}{c}{\textbf{PR Review Time}}\\ \hline
\multicolumn{2}{l|}{Adjusted $R^2$} & \multicolumn{2}{c}{0.2830}\\
\multicolumn{2}{l|}{Optimism-reduced adjusted $R^2$} & \multicolumn{2}{c}{0.2826}\\
\multicolumn{2}{l|}{Overall Wald $\chi^2$} & \multicolumn{2}{c}{17,690}\\ \hline
\multicolumn{2}{l|}{Spent Degrees of Freedom} & \multicolumn{2}{c}{24}\\ \hline
\multicolumn{2}{l|}{\textbf{Confounding Variables}} & Overall & Nonlinear\\ \hline
\multirow{2}{*}{PR Size} & D. F. & \multicolumn{2}{c}{\multirow{2}{*}{\dag}} \\
& $\chi^2$  & & \\ \hline
\multirow{2}{*}{\# Added lines} & D. F. & \multicolumn{2}{c}{\multirow{2}{*}{\dag}} \\
& $\chi^2$ &  & \\ \hline
\multirow{2}{*}{\#Deleted lines} & D. F. & 1 & \multirow{2}{*}{-} \\
& $\chi^2$  &  2.0$\ast$$\ast$$\ast$ &  \\ \hline
\multirow{2}{*}{Purpose} & D. F. & 2 & \multirow{2}{*}{-}  \\
& $\chi^2$ & 15.3$\ast$$\ast$$\ast$ &  \\ \hline
\multirow{2}{*}{\# Files} & D. F. & 1 & \multirow{2}{*}{-}\\
& $\chi^2$ & 1.5 &  \\ \hline
\multirow{2}{*}{Language}  & D. F. & 1 & \multirow{2}{*}{-}\\
& $\chi^2$ & 950.6$\ast$$\ast$$\ast$& \\ \hline
\multirow{2}{*}{\# Commits} & D. F. & 2 & 1 \\
& $\chi^2$ & 2960.0$\ast$$\ast$$\ast$  & 2950.5$\ast$$\ast$$\ast$  \\ \hline
\multirow{2}{*}{Description length} & D. F.  & 2 & 1 \\
& $\chi^2$     & 1338.4$\ast$$\ast$$\ast$                       & 954.0$\ast$$\ast$$\ast$                           \\ \hline
\multirow{2}{*}{PR author Exp.}             & D. F.    & 1                           & \multirow{2}{*}{-}            \\
                                            & $\chi^2$     & 46.9$\ast$$\ast$$\ast$                          &                               \\ \hline
\multirow{2}{*}{\# Comments}                & D. F.    & \multicolumn{2}{c}{\multirow{2}{*}{\dag}}                    \\
                                            & $\chi^2$     &                              &          \\ \hline
\multirow{2}{*}{\# Author comments}         & D. F.    & 2                           & 1                             \\
                                            & $\chi^2$     & 1518.4$\ast$$\ast$$\ast$                        & 790.0                           \\ \hline
\multirow{2}{*}{\# Reviewer comments}       & D. F.    & 3                           & 2                             \\
                                            & $\chi^2$     & 3349.9$\ast$$\ast$$\ast$                        & 2761.4$\ast$$\ast$$\ast$                          \\ \hline
\multirow{2}{*}{\# Reviewers}               & D. F.    & \multicolumn{2}{c}{\multirow{2}{*}{\dag}}                      \\
                                            & $\chi^2$     &                           &              \\ \hline
\multicolumn{2}{l|}{\textbf{Emoji Reaction Variables}} & Overall & Nonlinear \\ \hline
\multirow{2}{*}{With emoji reaction}        & D. F.    & \multicolumn{2}{c}{\multirow{2}{*}{\dag}}                     \\
                                            & $\chi^2$     &    &                                    \\ \hline
\multirow{2}{*}{\# Emoji reactions}         & D. F.    & 2                           & 1                \\
                                            & $\chi^2$     & 624.1$\ast$$\ast$$\ast$                        & 622.0$\ast$$\ast$$\ast$  
                                            \\ \hline
                                            
\end{tabular}
\begin{tablenotes}
      \scriptsize
      \item \dag: This explanatory variable is discarded during variable clustering analysis $\mid\rho\mid\geq$0.7
      \item -: This explanatory variable is not allocated with nonlinear degrees of freedom
      \item $\ast\ast\ast$: The explanatory power of this variable is statistically different, with p$<$0.001
\end{tablenotes}
\end{threeparttable}
}
\end{table}

\smallskip
\noindent
\textbf{\emph{Results.}} We now discuss the RQ1 results in the view of model construction and model analysis.

\textit{Model Construction.} Table~\ref{RQ1_Statistics} shows the model performance and statistics of the studied variables that are adopted in the regression model.
In the correlation and redundancy analysis, we remove those explanatory variables that are highly correlated with one another ($\rho$ value is greater than 0.7), i.e., PR size, \#Added lines, \# Comments, \# Reviewers, and With Emoji Reaction.
For the remaining explanatory variables, we do not find any redundant variable, i.e., the variable that has a fit with an $R^2$ greater than 0.9.
We then carefully allocated the degrees of freedom to the surviving variables, based on their potential for
sharing a nonlinear relationship with the dependent variable. 
As shown in the table, 24 degrees of freedom were spent on our constructed models.

\textit{Model Analysis.} We first examine the goodness of our model fit.
Table \ref{RQ1_Statistics} shows that the model achieves an adjusted $R^2$ score of 0.2830.
Similar to the prior work~\citep{Shopify_2018}, such an adjusted $R^2$ score is acceptable as our model is supposed to be explanatory not for the predictive purpose.
After applying the bootstrap validation approach, we observe that the optimism of an adjusted R2 is 0.0004, indicating that the constructed model does not have an overestimation issue and is stable to provide meaningful insight.

\begin{figure}[t]
    \centering
    \includegraphics[width=.8\linewidth]{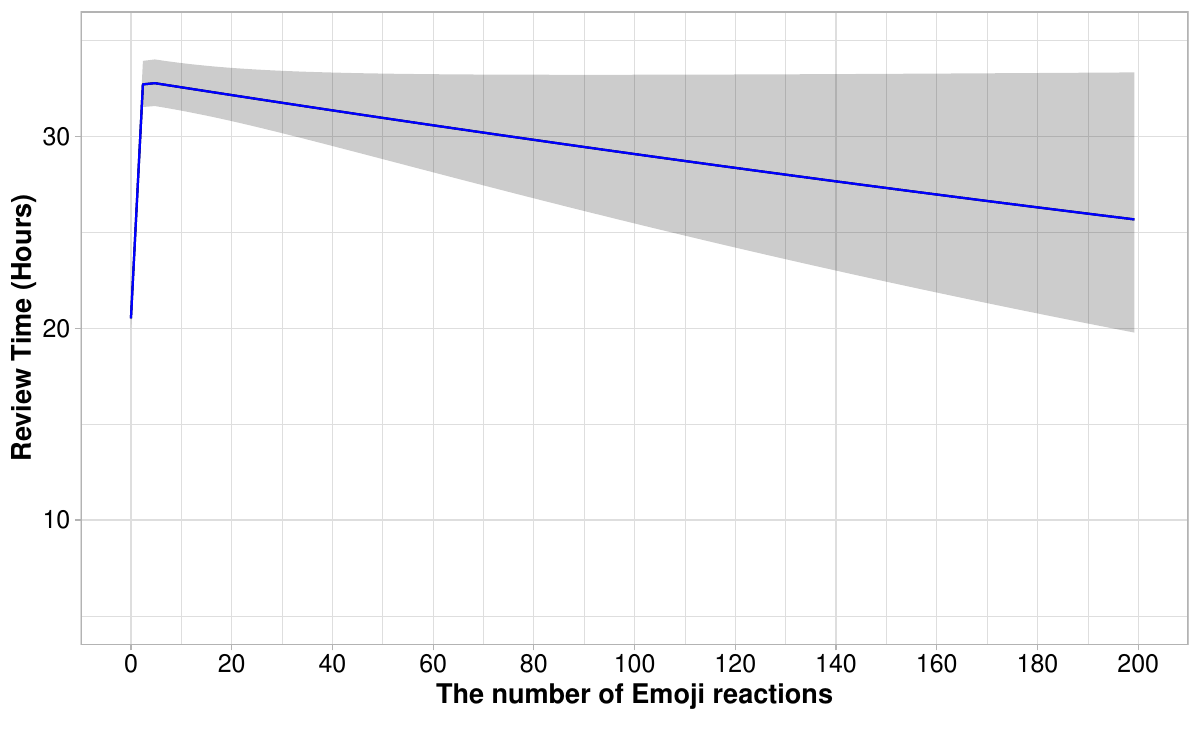}
    \caption{The nonlinear relationship between the likelihood that the review time is taken for a PR and the number of Emoji reactions (RQ1).}
    \label{fig:RQ1_likelyhood}
\end{figure}

We now discuss the explanatory power of the focused variable (i.e., Emoji Reaction Variables) and describe the relationship between this variable and the review time.
\textit{With Emoji Reactions} variable was removed due to its high correlation with another variable.
For the \textit{\# Emoji Reactions} variable, as shown in the table, we find that it has a significant correlation with the PR review time, with \textit{p-value $<$0.001}.
However, according to the Wald $\chi^2$ values, we observe that the explanatory power of the number of emoji reactions is not as large as the explanatory powers of the other dominant confounding variables.
The larger the $\chi^2$ of an explanatory variable is, the larger the contribution that the variable makes to the model.
Specifically, the Wald $\chi^2$ value of the \textit{\# Emoji Reactions} variable is 624.1, while the Wald $\chi^2$ values of the \textit{\# Reviewer Comments}, \textit{\# Commits}, \textit{\# Author Comments}, \textit{Description Length} are 3349.9, 2960.0, 1518.4, and 1338.4, respectively.
Figure ~\ref{fig:RQ1_likelyhood} depicts the nonlinear relationship between the number of emoji reactions and the review time.
As we can see, the number of emoji reactions shares a positive relationship when the number of emoji reactions is less than three.
But a negative relationship is observed if the number of emoji reactions is greater than three.
Based on the overall likelihood, the result suggests that compared to the PRs with no emoji reactions, it tends to take a relatively longer time for PRs with emoji reactions to be closed.

\begin{tcolorbox}[colback=gray!5,colframe=gray!75!black,title= RQ1: Does the emoji reaction used in the review discussion correlate with review time?]
Findings from the non-linear regression models indicate that the number of emoji reactions has a significant correlation with the review time.
In other words, PRs with emoji reactions are likely to have a longer review time when compared to those that do not.
\end{tcolorbox}

\subsection{First-time Contributors and Emoji Reactions (RQ2)}

\noindent
\textbf{\emph{Approach.}}To answer RQ2, we perform a quantitative analysis to investigate to what extent are PRs submitted by first-time contributors.
The first-time contributor in our study is defined as a contributor who has never submitted any PRs to the project.
Below we describe our approach in detail.

\textit{Proportion of PRs submitted by first-time contributors.} We use the dataset of PRs that contain at least one emoji reaction (i.e., 49,460 PRs as shown in Table~\ref{Data_summary}). 
We notice that a PR author is able to react emojis to comments. Since we focus on the emoji reactions that are received from developers (not the PR author), we first remove the emoji reactions that are from the PR author.
In the end, 36,902 PRs remained.
According to the explanatory variable (\texttt{\# PR author experience}) from RQ1, the dataset then is split into two groups: one group is labeled as the first-time contributor where the count of \texttt{\# PR author experience} is 0, the other group is labeled as the non first-time contributor where the count of \texttt{\# PR author experience} is greater than 0.
Afterward, we calculate the proportion of the PRs submitted by first-time contributors and non first-time contributors, respectively.

To validate the proposed hypothesis \textit{(H1): PRs submitted by first-time contributors receive more emoji reactions.}, we use the one proportion Z-test~\citep{z_test}.
One proportion Z-test compares an observed proportion to a theoretical one when
the categories are binary.

\begin{table}[b]
\centering
\caption{The proportion of PRs containing emoji reactions submitted by first-time contributors and non first-time contributors (RQ2).}
\label{tab:rq2_proportion}
\begin{tabular}{lrl}
\toprule
\textbf{PRs that contain emoji reactions} & \textbf{Percent (\%)} \\
\midrule
By first-time contributors &  3,855 (10.4\%) & \mybar{0.11}   \\
By non first-time contributors &  33,065 (89.6\%) &\mybar{0.89}   \\
\midrule
Total PRs &  36,920 & \\
\bottomrule
\end{tabular}
\end{table}

\smallskip
\noindent
\textbf{\emph{Results.}} Table ~\ref{tab:rq2_proportion} presents the proportion of PRs that contain emoji reactions (excluding the cases where the emojis are reacted by the PR authors) submitted by first-time contributors and non first-time contributors. 
As shown in the table, PRs submitted by non first-time contributors are more likely to receive emoji reactions, accounting for 89.6\% of the instances.
On the other hand, only 10.4\% of PRs submitted by first-time contributors received emoji reactions.

\textit{Significant Testing.} The statistical test (Z-test) reveals that there is a significant difference between the
proportion of PRs that receive emoji reactions submitted by the first-time contributors and the non first-time contributors, with a
p-value \textless 0.001.
This result indicates that the proposed hypothesis \textit{``(H1): PRs submitted by first-time contributors receive more emoji reactions.''} is not established.

\begin{tcolorbox}[colback=gray!5,colframe=gray!75!black,title= RQ2: Does a PR submitted by a first-time contributor receive more emoji reactions?]
A PR submitted by a first-time contributor is less likely to receive emoji reactions (i.e., 10.4\% of all PRs by first-contributors).
Statistically, the hypothesis that PRs submitted by first-time contributors receive more emoji reactions is not established.
\end{tcolorbox}

\subsection{Comment Intentions and Emoji Reactions (RQ3)}

\smallskip
\noindent
\textbf{\emph{Approach.}}
To answer RQ3, we conduct a quantitative analysis to investigate the PR comments that contain emoji reactions in two aspects: (I) the popularity of the comment intentions that contain emoji reactions, and (II) association mining between emoji reaction kinds, and between comment intentions and emoji reaction kinds.

\textit{Popularity of comment intentions with emoji reactions.} To categorize the intentions of the comments, we use the taxonomy of intentions proposed by \cite{huang2018automating}.
They manually categorized  5,408 sentences from issue reports of four projects on GitHub to generalize the linguistic pattern for category identification.
The definitions of intentions are described as follows:
\begin{itemize}
\item \textit{Information Giving (IG)}: Share knowledge and experience with other people, or inform other people about new plans/updates (e.g., ``The typeahead from Bootstrap v2 was removed.'').
\item \textit{Information Seeking (IS)}: Attempt to obtain information or help from other people (e.g., ``Are there any developers working on it?'').
\item \textit{Feature Request (FR)}: Require to improve existing features or implement new features (e.g., ``Please add a titled panel component to Twitter Bootstrap.'').
\item \textit{Solution Proposal (SP)}: Share possible solutions for discovered problems (e.g., ``I fixed this for UI Kit using the following CSS.'').
\item \textit{Problem Discovery (PD)}: Report bugs, or describe unexpected behaviors (e.g., ``the firstletter issue was causing a crash.'').
\item \textit{Aspect Evaluation (AE)}: Express opinions or evaluations on a specific aspect (e.g., ``I think BS3’s new theme looks good, it’s a little flat style.'').
\item \textit{Meaningless (ML)}: Sentences with little meaning or importance (e.g., ``Thanks for the feedback!'').
\end{itemize}
To facilitate the automation, they proposed a convolution neural network (CNN) based classifier with outstanding performance, which represents the state-of-the-art in the area of automatic intention mining.

\textit{Satiny Check.} To ensure that their automatic classifier is reliable enough to be used, we conducted a satiny check. 
To do so, we first randomly selected 30 comment samples and ran the automation to label these comments into the above seven intentions.
Then, the first two authors opened up a discussion to manually check whether the labeled intentions of the comments are correct or not, referring to the golden datasets from \cite{huang2018automating}.
After the check, we observe that 24 samples out of 30 samples (80\%) are correctly labeled by the automatic classifier.
We argue that this classifier is reliable enough to be used, since the average accuracy obtained in the original work in the context of issue comments is around 0.83.
Encouraged by the satiny check results, we will use the CNN based classifier to automatically label the intention of the comments that have emoji reactions.

After automatically labeling the intentions of comments, we group all the comments into seven intention categories and count their popularity.
Then, for each intention category, we calculate the occurrence of the emoji reaction kinds (i.e., \emojiThumbsUp, \emojiThumbsDown, \emojiGrinning, \emojiHeart, \emojiRocket, \emojiTada, \emojiEyes, \emojiConfused, and \emojiThumbsDown).
Note that when we calculate the occurrence, we did not take into account how many times an emoji reaction kind occurs in a comment.
For instance, supposing that a comment is reacted with THUMPS UP~\emojiThumbsUp~by five developers and HEART~\emojiHeart~by two developers, in this case, we count the occurrence of THUMPS UP~\emojiThumbsUp~and HEART~\emojiHeart~as one, respectively.

To validate the proposed hypothesis \textit{(H2): There
is a significant relationship between comment intentions and emoji reaction kinds}, we perform two significant tests.
First, we inspect whether or not the classified intentions of comments are normally distributed, using the Shapiro-Wilk test with alpha = 0.05~\citep{shapiro1965analysis}, a widely used normality test.
Second, for the relationship significance between comment intentions and emoji reactions, we use the Pearson's Chi-Squared test~\citep{pearson1900x}, which is commonly used for testing relationships between one or more categorical variables.

\textit{Association Mining.} To further explore the association (a) between emoji reaction kinds, and (b) between comment intentions and emoji reaction kinds, we leverage a rule-based machine learning method, widely known as the association rule technique~\citep{agrawal1993mining}.
Association rule mining is a technique to discover patterns of co-occurrence in a (large) dataset, by identifying entities that frequently appear together in a group, which is commonly used in SE community~\citep{prana2019categorizing, li2021understanding}.
To do so, we use the \texttt{association\_rules} function of the \texttt{mlxtend} Python package.

\begin{figure}[]
    \centering
    \includegraphics[width=.9\linewidth]{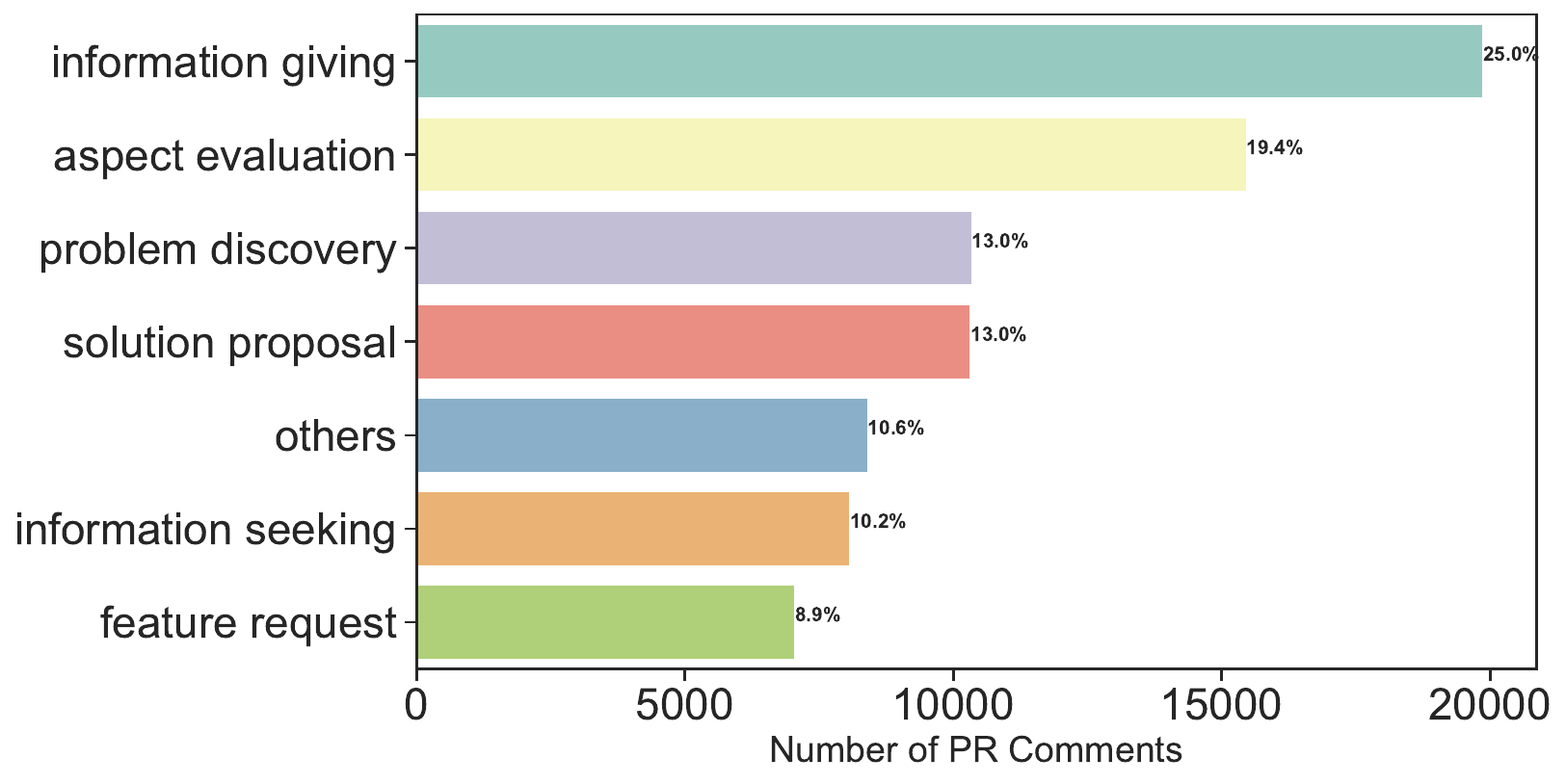}
    \caption{Intentions of the PR comments that have emoji reactions (RQ3).}
    \label{fig:prev_of_emoji}
\end{figure}

\smallskip
\noindent
\textbf{\emph{Results.}} 
\textit{Popularity of the comment intentions.} 
Figure \ref{fig:prev_of_emoji} presents the intention distribution of the PR comments that have emoji reactions.
As shown in the figure, we observe that the information giving category is the most common intention, 25.0\% being identified by the SOTA intention classifier.
This result indicates that developers tend to react to comments which share knowledge and experience with other people, or inform other people about new plans/updates, using emojis.
The second most frequent intention to receive emoji reactions is aspect evaluation (i.e., comments that express opinions or evaluations on a specific aspect), accounting for 19.4\%.
The least common intentions include information seeking and feature request, being 10.2\% and 8.9\%, respectively.

\begin{table}[]
\centering
\caption{The distribution of emoji reaction kinds across the comment intentions (RQ3).}
\resizebox{.9\textwidth}{!}{
\begin{tabular}{lrrrrrrrr}
\toprule
                    & \multicolumn{1}{c}{\emojiThumbsUp} & \multicolumn{1}{c}{\emojiHeart} & \multicolumn{1}{c}{\emojiTada} & \multicolumn{1}{c}{\emojiRocket} & \multicolumn{1}{c}{\emojiEyes} & \multicolumn{1}{c}{\emojiGrinning} & \multicolumn{1}{c}{\emojiThumbsDown} & \multicolumn{1}{c}{\emojiConfused} \\ \midrule
Information Giving  & 63.6\%                        & 11.2\%                    & 10.7\%                     & \textbf{\textcolor{blue}{6.3\%}}                      & 2.6\%                    & 3.0\%                     & 1.4\%                            & 1.0\%                        \\
Information Seeking & 70.6\%                        & 9.2\%                     & 6.5\%                      & 3.8\%                      & 3.4\%                    & 4.6\%                   & 1.0\%                            & 0.8\%                        \\
Aspect Evaluation   & \textbf{\textcolor{blue}{74.4\%}}                      & 7.7\%                     & 5.7\%                      & 3.5\%                      & 2.4\%                    & 3.5\%                     & \textbf{\textcolor{blue}{1.5\%}}                            & 1.1\%                        \\
Feature Request     & 68.6\%                        & 10.8\%                    & 8.0\%                      & 5.4\%                      & 2.7\%                    & 2.7\%                     & 1.2\%                            & 0.8\%                        \\
Problem Discovery   & 67.1\%                        & 9.8\%                     & 8.6\%                      & 5.3\%                      & \textbf{\textcolor{blue}{3.9\%}}                    & 2.7\%                     & 1.1\%                            & \textbf{\textcolor{blue}{1.3\%}}                        \\
Solution Proposal   & 67.6\%                        & 10.4\%                    & 9.1\%                      & 5.2\%                      & 2.4\%                    & 3.0\%                     & 1.2\%                            & 0.9\%                        \\
Others              & 58.4\%                        & \textbf{\textcolor{blue}{14.5\%}}                    & \textbf{\textcolor{blue}{13.9\%}}                     & 5.1\%                      & 1.5\%                    & \textbf{\textcolor{blue}{5.5\%}}                     & 0.5\%                            & 0.5\%                        \\ \bottomrule
\end{tabular}}
\label{results:rq3_emoji}
\end{table}

\begin{table}[]
\centering
\caption{Association rules at the emoji reaction level (RQ3).}
\label{rq3:asso_emoji}
\begin{tabular}{lrr}
\toprule
Rule & Support & Confidence \\ \midrule
\{\emojiEyes (\texttt{EYES}), \emojiHeart (\texttt{HEART}), \emojiRocket (\texttt{ROCKET})\} $\Rightarrow$ \{\emojiTada (\texttt{HOORAY})\}      & 0.0017  & 0.86       \\
\{\emojiEyes (\texttt{EYES}), \emojiHeart (\texttt{HEART}), \emojiRocket (\texttt{ROCKET})\} $\Rightarrow$  \{\emojiThumbsUp (\texttt{THUMBS UP})\}    & 0.0016  & 0.82       \\
\{\emojiEyes (\texttt{EYES}), \emojiHeart (\texttt{HEART}), \emojiTada (\texttt{HOORAY})\} $\Rightarrow$  \{\emojiThumbsUp (\texttt{THUMBS UP})\}    & 0.0016  & 0.82       \\
\bottomrule
\end{tabular}
\end{table}

\begin{table}[]
\centering
\caption{Association rules at the intention level (RQ3).}
\label{rq3:asso_intention}
\begin{tabular}{lrr}
\toprule
Rule & Support & Confidence \\ \midrule
\{aspect evaluation\} $\Rightarrow$  \{\emojiThumbsUp (\texttt{THUMBS UP})\}     & 0.16  & 0.80       \\
\{information giving\} $\Rightarrow$  \{\emojiThumbsUp (\texttt{THUMBS UP})\}    & 0.18  & 0.72       \\
\bottomrule
\end{tabular}
\end{table}

To complement the insights, we also investigate how frequently each kind of emoji reaction is used across the eight comment intentions.
Table~\ref{results:rq3_emoji} shows the related results.
In general, we observe that positive emoji reactions are frequently used (\emojiThumbsUp, \emojiHeart, \emojiTada) when compared to negative emoji reactions (\emojiThumbsDown, \emojiConfused).
For instance, the usage frequency of THUMPS UP~\emojiThumbsUp~on average reaches up to 67.2\% across seven intentions, while the usage frequency of THUMPS DOWN~\emojiThumbsDown~on average is only 0.9\%.
Upon a look at the emoji reaction kinds, we find that each emoji reaction kind tends to be used with specific intentions.
For THUMPS UP~\emojiThumbsUp, as highlighted in blue, it is most frequently used in the aspect evaluation intention (74.4\%). 
For HEART~\emojiHeart, HOORAY~\emojiTada, and LAUGH~\emojiGrinning, they are likely to be used in Others intention (sentences with little meaning or importance, e.g., ``Thanks for the feedback!''), i.e., 14.5\%, 13.9\%, and 5.5\%, respectively.
The EYES~\emojiEyes~ is commonly used in problem discovery intention (3.9\%).

\textit{Significant Testing.} First of all, the Shapiro-Wilk test suggests that the classified intentions of comments are normally distributed, with alpha = 0.12 (greater than 0.05).
The Pearson's Chi-Squared test confirms that there is a significant relationship between comment intentions and emoji reactions, i.e., p-value \textless 0.001, indicating that the hypothesis (H2) is established.

\textit{Association mining.} We apply the association rule mining at two levels: between emoji reaction kinds, and between emoji reaction kinds and comment intentions.
With regard to the extracted rules at the emoji reaction level, we only consider rules
with the support of at least 0.0013 (i.e., the rule must apply to at least six
emoji reaction kinds) and the confidence of at least 0.8, similar to the prior work~\citep{prana2019categorizing}. 
Table~\ref{rq3:asso_emoji} shows the three extracted association rules at the emoji reaction level.
For example, the comment is reacted by the group of \{\emojiEyes, \emojiHeart, \emojiRocket\} is likely to contain \emojiTada, with the confidence of 0.86.
In terms of the intention level, we set the support of at least 0.15 and the confidence of at least 0.7.
Table \ref{rq3:asso_intention} shows the two extracted association rules at the intention level.
We observe that the comment intentions of aspect evaluation or information giving are likely to be reacted with \emojiThumbsUp~(\texttt{THUMBS UP}), the confidence being 0.80 and 0.72, separately.

\begin{tcolorbox}[colback=gray!5,colframe=gray!75!black,title= RQ3: What is the relationship between the intention of comments and their emoji reaction?]
We find that PR comments with the intention of \textit{information giving} are more likely to receive emoji reactions (i.e., 25.0\%). Our statistical test suggests that there is a significant relationship between comment intentions and emoji reactions.
Moreover, the positive THUMBS UP (67.2\% on average) is widely used, while negative emojis (between 0.5\% to 1.5\%) are rarely used.
\end{tcolorbox}

\subsection{Consistency of Emoji Sentiments (RQ4)}
\noindent
\textbf{\emph{Approach.}}
To answer RQ4, we conduct a mixed analysis (qualitative and quantitative) to investigate the PR comments that contain emoji reactions in terms of (I) frequency of sentiment consistency between emoji reactions and comments, and (II) reasons behind the inconsistency.
Below, we describe the approach of two aspects in detail.

\textit{Frequency of sentiment consistency.} To investigate the sentiment consistency, first we need to label the sentiment of emoji reactions and the sentiment of comments, separately.
\textit{For the sentiment of emoji reactions}, we refer to our preliminary study in the registered report and we classify the sentiments of the emoji reactions into the following four types: \texttt{Positive}, \texttt{Negative}, \texttt{Neutral}, and \texttt{Mixed}.
\texttt{Positive} refers to the single usage or the combination usage of \texttt{THUMBS UP} \emojiThumbsUp , \texttt{LAUGH} \emojiGrinning, \texttt{HOORAY} \emojiTada, \texttt{HEART} \emojiHeart, and \texttt{ROCKET} \emojiRocket~reactions.
\texttt{Negative} denotes to the single usage or the combination usage of \texttt{THUMBS DOWN} \emojiThumbsDown,and \texttt{CONFUSED} \emojiConfused~reactions.
\texttt{Neutral} represents the usage of \texttt{EYES} \emojiEyes~reaction.
\texttt{Mixed} refers to the combination usage of the four categories mentioned above.
\textit{For the sentiment of comments}, we use SentiStrength-SE~\citep{Islam2018SentiStrengthSEED}, a state-of-the-art sentiment analysis tool that utilizes domain dictionary and heuristics for software engineering text.
In our study, input is the PR comment that contains emoji reactions and output is a sentiment score ranging from -5 (very negative) to 5 (very positive).
Note that, to reduce the potential threat due to false positives, we exclude the PR descriptions that contain emoji reactions.
After we obtain the sentiment labels for the emoji reactions and the comments, we then map them into the level of a PR comment and count the frequency of the possible patterns (e.g., Positive-Positive, Positive-Neutral, Positive-Negative, and so on). 

To validate the proposed hypothesis \textit{(H3): There is a significant relationship between comment sentiments and emoji reaction sentiments}, similar to RQ3, we perform the Pearson's Chi-Squared test to confirm if a significant relationship exists or not.

\textit{Reasons behind the inconsistency.} To investigate the reasons of the inconsistency between the sentiment of emoji reaction and PR comment, we perform a manual coding. 
In this study, we focus on the inconsistency in terms of the positive and negative aspects.
It refers to the case where the sentiment of the emoji reaction is positive while the sentiment of the comment is negative, or vice versa.
Since there is no specifically available reason taxonomy to refer to, we apply an open coding approach~\citep{charmaz2014constructing} to classify randomly sampled comments in inconsistent cases.
To discover as complete of a list of reasons as possible, we strive for \textit{theoretical saturation}~\citep{eisenhardt1989building} to achieve analytical generalization.
Similar to the prior work~\citep{xiao2021characterizing}, we initially set our saturation criterion to 50.
Then the first two authors continue to code randomly selected inconsistent comments until no new codes have been discovered for 50 consecutive comments.
If the new codes occur, we performed
another pass over all of the comments to correct miscoded
entries and tested the level of agreement of our constructed codes, since codes that emerge late
in the process may apply to earlier reviews.
The third author joins the open discussion when disagreements occur and validates the suggested codes.
Finally, we reach saturation after coding 100 samples.

\begin{figure}[t]
    \centering
    \includegraphics[width=.7\linewidth]{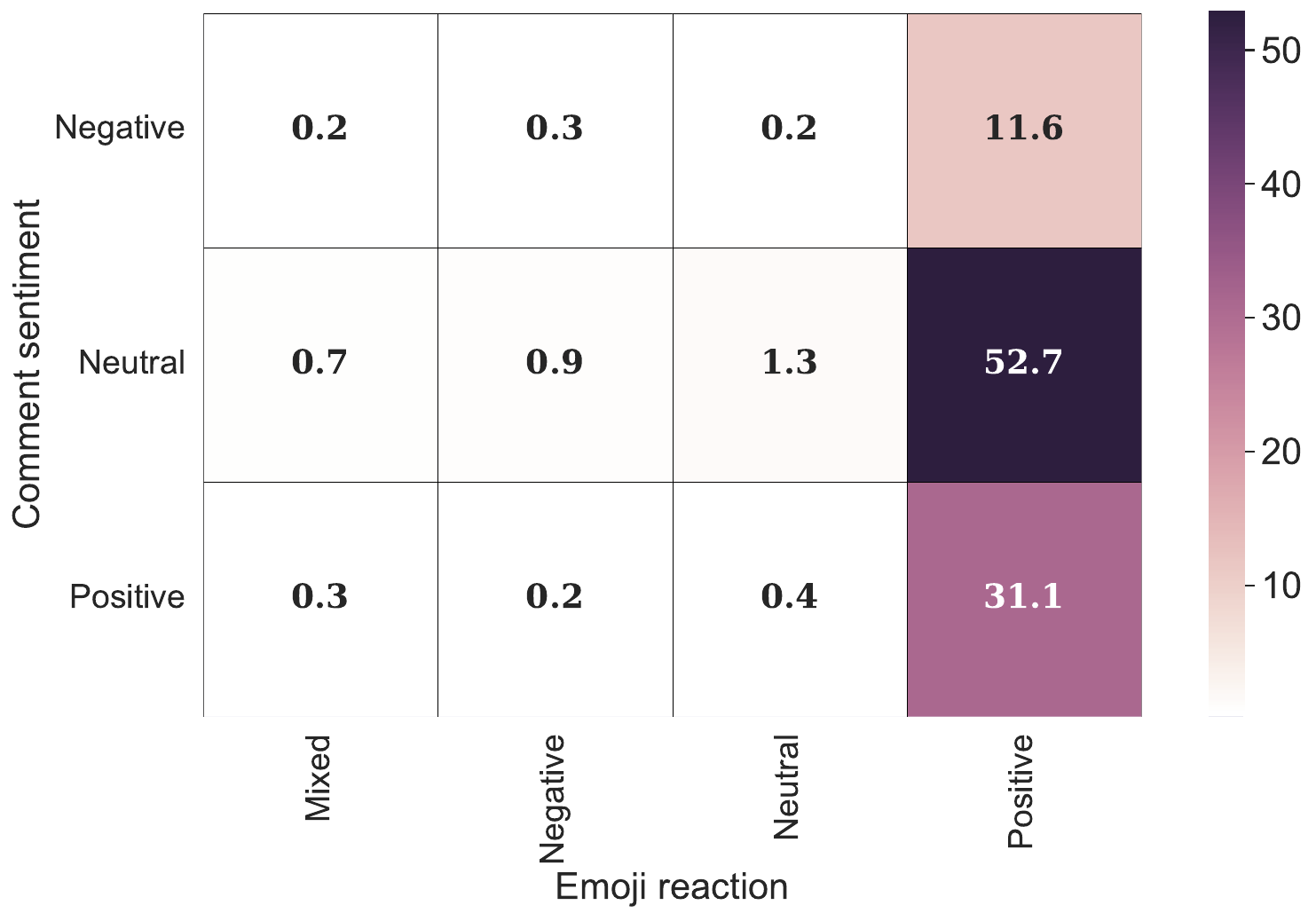}
    \caption{Pair-wise between emoji reactions and comment sentiments (RQ4). The annotations in the plot are in the form of \%.}
    \label{fig:rq4_heatmap}
\end{figure}

\smallskip
\noindent
\textbf{\emph{Results.}}
\textit{Frequency of sentiment consistency}. Figure~\ref{fig:rq4_heatmap} shows the pair-wise distribution between the sentiments of emoji reactions and comments.
As shown in the figure, we observe that the major pair-wise is Positive-Neutral (i.e., Positive emoji reaction and Neutral comment), accounting for 52.7\%.
The second most frequent pair-wise is Positive-Positive, being 31.1\%.
At the same time, we find that the proportion of Positive-Negative (i.e., Positive emoji reaction and Negative comment) pair-wise is not high as expected, 11.6\% being identified, ranked as the third most common pair-wise.
Conversely, the Negative-Positive (i.e., Negative emoji reaction and Positive comment) accounts for only 0.2\%.

\textit{Significant Testing.} The Pearson's Chi-Squared test suggests that there is a significant relationship between comment sentiments and emoji reaction sentiments, i.e., p-value \textless 0.001, indicating that our proposed hypothesis (H3) is established.

\textit{Reasons behind sentiment inconsistency}. Nine reasons are identified from our open coding process (i.e., 100 samples).
Table \ref{rq4:inconsistent} presents the frequency of these reasons, with their representative samples.
We find that \textit{Acknowledge a mistake} is the most common reason behind sentiment inconsistency, accounting for 22\%.
The following reason is \textit{Acknowledge a proposal}, being 15\% classified.
The rare reasons include \textit{Irony} and \textit{Disagree with optimistic proposal} (i.e., 1\% for them, respectively).
On the other hand, we observe that the rate of False-Positive is relatively high, accounting for 23\%.
False-Positive refers to those cases, where the sentiments of comments are not correctly labeled by the tool after the manual validation.

\begin{table}[]
\centering
\caption{Frequency of reasons behind sentiment inconsistency (RQ4).}
\label{rq4:inconsistent}
\resizebox{.99\columnwidth}{!}{
\begin{tabular}{l|p{6cm}rr}
\toprule
\textbf{Reason} & \multicolumn{2}{c}{\textbf{Representative example}}  & \textbf{\# Cases}  \\ \midrule
Acknowledge a mistake  & ``Oops - pushed to wrong branch... undoing now!!'' & \emojiGrinning & 22       \\
Acknowledge a proposal   & ``I don't really like this solution. The join will not correctly set-up the model graph, so the best solution in my opinion is to write: @code...''  & \emojiThumbsUp & 15    \\
Counter a pessimistic opinion   & ``Faker is a dev dependency, and would fail to run in production.
I afraid that developer may start using this method beyond seeders''  & \emojiThumbsUp & 13      \\
Confirm a proposal   & ``Sorry about the last minute change, but I realized that client-side could also benefit from richer exception messages. I'd appreciate if reviewers can take a look at the following commit 58acdbc''  & \emojiThumbsUp &11       \\
Acknowledge a pessimistic outcome    & ``Sorry @username - this is way out of the areas of anything I have knowledge of''  & \emojiThumbsUp & 8       \\
Confirm request for a pessimistic outcome  & ``Thanks for the PR!
So, the change is that: @code...
but this is probably there to prevent bad things to happen like a sql injection, no?
PS: I can't easily read the diff on github, dunno why, if some other staff member can explain why, it's very annoying (seems related to codacy)'' & \emojiThumbsUp & 3       \\
Counter an optimistic proposal   & ``@username... How good a reason is depends not only on the severity of the issue but also on the importance of the data we update or not update. @explanation...''  & \emojiThumbsDown & 1      \\
Irony    & ``Worst Translation Ever''  & \emojiThumbsDown & 1       \\
Others   &   & & 3       \\
False-Positive    & Comment sentiments are not correctly labeled  by the tool. & & 23       \\
\bottomrule
\end{tabular}}
\end{table}


\begin{tcolorbox}[colback=gray!5,colframe=gray!75!black,title= RQ4: Is emoji reaction consistent with comment sentiment?]
Results indicate that a high percent of emoji sentiments are consistent with their comments, with only 11.8\% (i.e., 11.6\% + 0.2\%) being inconsistent.
The statistical test confirms that there is a significant relationship between comment sentiments and emoji reaction sentiments.
We then identify nine reasons behind these inconsistencies (acknowledgments, confirmations, counters, irony, etc.).
A positive acknowledgment of a negative mistake is the most frequent reason (i.e., 22\%).
\end{tcolorbox}

\section{Discussion}
We now discuss the implications of our empirical findings, provide possible suggestions to facilitate code review and social communication, and outline the limitations and potential research topics. 

\paragraph{\textbf{Implications for project development on GitHub.}} Developers on GitHub should be aware that using an emoji reaction is correlated with the review process. 
Our RQ1 results indicate that emoji reactions have a statistical correlation with review time, and PRs with emoji reactions overall tend to take a longer review time than the ones with no emoji reactions.
It is important to note that we do not claim causality, as there could be other confounding factors that play a role in the regression models.
We speculate that PRs with emoji reactions are likely to include complex contexts that require further discussions and may be involved more participants compared to the ones that do not contain emojis.
Usually, simple PRs are handled quickly, even when they do not have reactions.
Hence, such a finding does not imply that emoji reactions have a negative effect on review efficiency as the efficiency could be affected by the nature of PRs. 
Potential reasoning may be related to the findings observed in the context of issue reports.
Prior work~\citep{borges2019beyond} reported that issues with reactions usually take more time to be handled and have longer discussions, especially for those complex bugs and enhancements.
Furthermore, due to longer discussions, cognitive loads combined with textual and visual expressions may increase.
For example, \citet{tigwell2016oh, bai2019systematic} pointed out that the difference in how emoji are understood could bring ambiguities in the interpretation of communication, and leads to inefficiency.

The usage of emoji reactions signals an already friendly environment on GitHub.
This is evident by RQ3, where positive emoji reactions are commonly used during PR comments across different intentions, especially THUMBS UP being 67.2\% on average (Table 5). 
Negative emoji reactions are rarely used, only ranging from 0.5\% to 1.5\%.
Our RQ3 results suggest that contributors tend to react to those comments with positive emojis that share knowledge and require opinion expressions or evaluations.
Specifically, We find that those intentions of information giving and aspect evaluation are more likely to receive emoji reactions (25.0\% and 19.4\%, respectively, shown in Figure 3). 

Emoji reactions on GitHub have the potential to reduce toxicity.
During the reason analysis behind the sentiment inconsistency (RQ4), results show that in most cases, emoji reactions are used as a kind of encouragement to react to negative comments.
Even in cases where the reaction is not consistent with the sentiment of the comment, as shown in Table 8, the most prevalent reason is to acknowledge a mistake by replying to a negative comment with a positive emoji reaction.
This indicates that emoji reactions can reduce or counter negative sentiments, and potentially lighten tense situations. 
The negative effect of toxicity in open-source projects is widely studied.
Literature shows evidence that some newcomers disengage due to negative interactions (Qiu et al., 2019), and frequent contributors often suffer from stress and burnout facing toxic interactions (Raman et al., 2020).
Along the same line to support social communication, our study results prove the positive role of emoji reactions play in the collaboration environment.

\paragraph{\textbf{Suggestions.}} We provide several possible suggestions for the stakeholders.
On the one hand, contributors should not be afraid of expressing negative or contentious comments, as emoji reactions have the potential to diffuse any toxicity and can lighten the tense situation.
Especially for newcomers, they should not expect immediate emoji reactions as a sign of hostility. 
Instead, emoji reactions could be considered an indicator of familiarity and positiveness in the environment.
On the other hand, in addition to diffusing tense situations, reviewers are encouraged to keep using emoji reactions to properly express their sentiments in order to construct a friendly and positive interaction with review participants. 
In terms of governance and maintaining consistency, GitHub projects are encouraged to document, broadcast, and mentor the proper usage of emoji reactions for the sustained livelihood of Open Source projects.

\paragraph{\textbf{Limitations
and potential research topics.}} Our research foresees research directions for code review efficiency, the health of communication channels, and the attraction, onboarding, and sustainability of contributions to Open Source Software.
To address the limitation of the statistical models, future work should further investigate the causality of the  relationship between emoji reactions and review time to understand whether or not the longer time has some negative or positive effects on the project. 
RQ2 results show that non first-time contributors are more likely to receive emoji reactions, hence future work could further explore what features will contribute to this.
Several studies demonstrated that the usage of emojis could be affected by developer gender, age, and cultural difference~\citep{guntuku2019studying, herring2020gender}. 
Therefore, another potential direction is to look into whether such factors play a role in the usage of emojis within the scope of code review.


\section{Threats to Validity}
In this section, we disclose the threats to the validity of our study.

\paragraph{\textbf{External validity.}} This validity refers to the generalization ability of our results.
We conduct an empirical study on 1,850 GitHub repositories across seven popular languages. 
The threat may occur due to the number of studied repositories and the choice of languages.
To relieve this threat, we construct a representative dataset, with a confidence level of 95\% and a confidence interval of 5, by taking each language-based repository population into account.
Meanwhile, these seven popular languages are commonly studied in the prior study~\citep{hata}.
We believe that our study on these 1,850 repositories is sufficient enough to shed light on the role of the emoji reactions in the code review process.

\paragraph{\textbf{Construct validity.}} This validity denotes the degree to which our measurements capture.
We summarize four threats.
The first threat may exist in the bot removal in the data preparation process.
We rely on the combination of two bot detection techniques (``bot'' suffix and list of ground-truth bots).
It is reasonable that we can not ensure we are able to remove all the GitHub bots.
However, the highest accuracy of these two bot detection techniques is recognized in the latest research. 
Second, identifying the purpose of PRs in RQ1  may introduce a potential threat. 
The purposes in fact would not only include documentation, bug fixing, and feature introduction but also include others like refactoring activity.
However, there is no reliable method to automatically identify this activity.
Hence, along the same line as the existing work~~\citep{EMSE2017_Pick, shane2014msr, WangEMSE2021}, we decided to adopt the same automated identification method.
Third, other diversity-related factors (e.g., gender, age, and culture differences) could potentially influence the findings on the usage of emoji reactions in RQ3 and RQ4.
Thus, future work should further investigate the effect of these factors in the context of code review.
The last threat could exist in our RQ4 qualitative analysis to categorize the reasons behind the sentiment inconsistency.
To mitigate this threat, we performed an open coding to ensure all the potential codes, carefully following the guidelines provided by the literature. 
For instance, we reached saturation until no new codes emerged (i.e., 100 samples).

\paragraph{\textbf{Internal validity.}} Internal validity refers to the approximate truth about inferences.
Two potential threats are concluded.
The first threat may occur, resulting from the selected classifier or tool.
In RQ3, to label the comment intention, we use the automatic classifier, which is considered as SOTA in the context of intention mining. 
The comment intention could be mislabeled due to the classifier uncertainty.
Thus, to mitigate this threat, we performed a satiny check on 30 samples and the accuracy was promising, i.e., 80\%.
In RQ4, the comment sentiment could as well be mislabeled since we rely on the SentiStrength-SE tool.
However, the SentiStrength-SE tool is recognized as one of the SOTA sentiment tools and is widely adopted in the SE domain.
The second threat could exist in the selection of statistical significance testing techniques.
To validate the establishment of our three proposed hypotheses, we apply several kinds of significance testing techniques (i.e., Z-test, Pearson's Chi-Squared test). 
The casual-effect may vary on the selection of these techniques.
We are however confident, as these techniques are commonly used in empirical studies.

\section{Related Work}
This section situates this study with respect to the related work, including the effect of non-technical factors in code review, and modern communication in software development.

\subsection{Effect of Non-technical Factors in Code Review}
Software code review is established as one of the best practices to improve software quality~\citep{Rigby_2011}. 
Tool-based code reviews, a lightweight variation of traditional code reviews, have been widely studied in the last decade.
A number of studies point out that the effect of tool-based review is not only affected by technical factors, but also by non-technical factors~\citep{s32, Olga_2016}.
In terms of developer participation, \cite{shane2014msr} reported that developer participation in code review is also associated with the incidence of post-release defects.
Similarly, \cite{kononenko2015investigating} found that personal metrics and participation metrics shared a relationship with the quality of the review process.
Review comment is one of the main building blocks and is crucial to the review quality~\citep{sadowski2018modern}.
\cite{jiang2013will} showed that the amount of
discussion is a significant indicator of whether a patch will be accepted for integration into the Linux kernel.
\citet{el2019empirical} empirically studied the impact of sentiment embodied within developers’ comments and they observed that the reviews with negative comments on average took more time to complete than the reviews with positive/neutral comments.
\cite{hirao2020code} studied divergent review scores in the review discussion and suggested that divisive patches are often integrated through discussion, integration timing, and careful revision.
\cite{zhang2022pull} conducted an empirical study to systematically investigate the factors that affect pull request latency.
They found that comments have a significant impact on the latency of pull requests.

With the wider usage of emoji reactions in the code review (i.e., pull requests), we argue that as a non-technical factor, emoji reactions may also have an effect on the review process.

\subsection{Modern Communication in Software Development}
For purposes of collaboration and communication between developers, communication channels (e.g., issue reports, mailing lists, pull requests, and GitHub discussions) are integrated or supplemented in development tools~\citep{chui2012social, tantisuwankul2019topological}. 
Various non-textual information is embedded in these communication channels to enrich the knowledge sharing between developers. 
\cite{nayebi2020eye} showed that the increasing trend of image usage and images facilitated Q\&A communication in Stack Overflow. 
Recent studies also investigate the usage of these visual contents in terms of issue reports \citep{agrawal2022understanding, kuramoto2022visual}.
Link sharing is also another common modern communication strategy.
\cite{hata} investigated the characteristics 
and phenomenons of links in source code comments and found that referencing links is prevalent in source code comments (e.g., software homepages, and articles or tutorials).
\cite{WangEMSE2021} observed that the practice of link sharing has a significant correlation with the code review time. 
\cite{fu2022understanding} found that code snippets are not frequently used in code review, however, code snippets with the aim of suggestion are actively accepted by the patch author. 

Developers also use emojis, visual symbols in computer mediated communication, to represent their opinions~\citep{bai2019systematic}. 
The prior work has shown that emoji use in
the East and the West reveals recognizable normative and culture specific patterns~\citep{guntuku2019studying}.
\cite{claes2018use} also observed that various types of emojis can be used in two studied issues trackers and from different developers (i.e., western developers use more emojis than eastern developers) at different times (e.g., negative emojis are used during weekends).
\citet{herring2020gender} pointed out that females use emoji and emoticons
more frequently than males do.
\cite{borges2019beyond} explored the usage of GitHub reactions in issues reports.
They found that the usage of emoji reactions is increasingly growing and furthermore emoji reactions make more discussion for bug and enhancement issue reports.
To support visual representations of sentiments, \citet{venigalla2021stackemo} proposed a plugin namely StackEmo to augment comments on Stack Overflow with emojis.
\citet{chen2021emoji} learned representations of SE-related texts through emoji prediction by leveraging Tweets and GitHub posts containing emojis.
\cite{rong2022empirical} discovered the usage of emojis in software development.
Their results showed that emojis tend to be used during every phase of a conversation on GitHub regardless of the developers' roles.

Although the phenomenon of emoji usage in textual content has been commonly addressed, it is still unclear what is the role of the emoji as a reaction during the software development (i.e., code review process). Our study would complement the knowledge lying in the context of modern communication.

\section{Deviations from the Registered Report}
Our study requires unavoidable deviations from the research protocol~\citep{son2021more} that
we have carefully documented below:
\begin{enumerate}
    \item \textit{Studied Repository Dataset.}
    Due to the restriction of GitHub API downloading within our given time-frame, we decided to select a representative sample of the studied repositories instead of the full dataset.
    Hence, we constructed a representative repository dataset that contains 1,850 repositories across seven languages, instead of the 25,925 repositories outlined in the protocol. As described in the data collection (Section 3), this is a statistical sample that was systematically collected. 
    \item \textit{Research Approach.} We outline three deviations related to the methodology, and have a minimal effect on the results. First, in the explanatory variable selection (RQ1), we added another two variables (i.e., Language and Description length) that show the effect in the prior work. 
    Meanwhile, we removed the variable ``commit size'' and instead we introduced another two related variables ``\# Added lines'' and ``\#Deleted lines''. In addition, we altered the calculation of the dependent variable ``Review time'' by taking into account the PR closed time.
    Second, in the proportion of PRs submitted by the first-time contributors (RQ2), we did not follow a control study where we planned to construct a balanced control group. Instead, we divided all PRs that contain emoji reactions into the ones by first-time contributors and the other ones by non first-time contributors.
    Third, during the manual classification of reasons behind sentiment inconsistency (RQ3), we did not calculate the Kappa score as the open coding process does not require it~\citep{hirao2019review, xiao2021characterizing}.
    \item \textit{Hypothesis Testing.} While conducting the experiments, we realized that the hypotheses and corresponding statistical tests reported in the protocol were not appropriate from RQs2--4. For RQ2, the result was in the form of binary categories, thus we changed to use the one proportion Z-test. For RQ3, we changed the prior hypothesis (H2) to the hypothesis ``There is a significant relationship between comment intentions and emoji reaction kinds''.
    To validate it, we used the Pearson's Chi-Squared test. 
    For RQ4, we changed the hypothesis (H3) to the hypothesis ``There is a significant relationship between comment sentiments and emoji reaction sentiments''.
    Similarly, we adopted the Pearson's Chi-Squared test to validate it.
\end{enumerate}

\section{Conclusion}
In this work, we conducted an empirical study on 1,850 repositories to investigate the role of emoji reactions in GitHub pull requests.
Specifically, we analyzed the following four aspects: (i) the correlation between the emoji reactions and the review time, (ii) whether first-time contributors being likely to receive emoji reactions,
(iii) relationship between the comment intentions and emoji reactions, and (iv) consistency between comment sentiments and emoji reaction sentiments.

The results show that (i) the number of emoji
reactions have a significant correlation with the review time; (ii) a PR submitted by a first-time contributor is less likely to receive emoji reactions; (iii) the PR comments with the intention of information giving are more likely to receive emoji reactions; and (iv) Positive–Negative inconsistency pair-wise accounts for 11.8\%, and to acknowledge a mistake is the most common reason of sentiment inconsistency.
These empirical results highlight the role of emoji reactions play in collaborative communication and specifically suggest that the usage of emoji reactions signals an already positive environment on GitHub and it has the potential to reduce toxicity. 
Future research directions include a deeper
study of investigating the causality of emoji reactions and understanding the reasons
why it takes a longer time to complete the review, and analyzing the diversity-related perspectives of emoji reactions in the scope of code review.


\section*{}

\textbf{Acknowledgements}
This work is supported by Japanese Society for the Promotion of Science (JSPS) KAKENHI Grant Numbers 18H04094 and 20K19774 and 20H05706.
\\
\\
\noindent \textbf{Data Availability}
The datasets generated during and/or analysed during the current study are available in the GitHub repository, \url{https://github.com/NAIST-SE/EmojiReaction_PR}.

\section*{Declarations}
\textbf{Conflict of Interests}
The authors declare that Raula Gaikovina Kula and Yasutaka Kamei are members of the EMSE Editorial Board. 
All co-authors have seen and agree with the contents of the manuscript and there is no financial interest to report.

\bibliographystyle{spbasic}
\bibliography{myref}

\end{document}